\documentclass[doublecol]{epl2}

\title{Testing the special relativity theory with neutrino interactions}

\author{P.W. Cattaneo \inst{1} }

\institute{
\inst{1} INFN, Sezione di Pavia - Via Bassi 6, I-27100, Pavia, Italy 
}

\date{\Today}


\abstract{
A recent measurement of neutrino velocity by the OPERA experiment 
and prediction of energy loss of superluminal 
neutrino via the pair creation process $\nu\to \nu e^+e^-$ stimulated
a search of isolated $e^+e^-$ pairs in detectors 
with good tracking capability traversed by a large flux of high energy neutrino 
like NOMAD. NOMAD has already searched for similar topologies. These results 
can be reinterpreted to provide stringent 
limits on special relativity violating parameters separately for each $\nu$ species.
}

\begin{document}

\maketitle

\section{Introduction}

The measurement of $\nu$ velocity recently \cite{opera} reported by 
the OPERA experiment stimulated theoretical studies \cite{glashow} (see also
\cite{bi,mestres}) about the consequence of $\nu$ 
superluminality \footnote{Results presented at Neutrino 2012 conference in Kyoto, Japan 
by the OPERA, LVD, ICARUS, Borexino, T2K and MINOS collaborations (see also \cite{icavel})
rule out the original OPERA 
claim and are consistent with the neutrinos moving at the speed of light within $\sim\ 10^{-6}$.}.

They predict that that a superluminal $\nu$ 
would loose energy
through Cerenkov-like processes like

\begin{equation}
\nu_\mu \rightarrow \left\{ \begin{array} {ll}
\nu_\mu + \gamma & (a) \\
\nu_\mu + \nu_e +\overline \nu_e & (b) \\
\nu_\mu + e^+ + e^- & (c) 
\end{array} \right .
\label{nuee}
\end{equation}

The main effective process of energy loss is expected to be (c), that has a threshold 
$E_0=2m_ec^2/\sqrt{\delta_\nu}$, where $m_ec^2$ is the electron rest energy and 
$\delta_\nu=\beta^2_\nu - 1 \approx 2\times (\beta_\nu - 1)$. 
\cite{glashow,bi} report that the rate of pair emission and rate of energy loss are
\begin{eqnarray}
\Gamma &=& k^\prime \frac{G^2_F}{192\pi^3}E_\nu^5\delta_\nu^3 \\
\frac{dE}{dx} &=& -k \frac{G^2_F}{192\pi^3}E_\nu^6\delta_\nu^3 
\label{nudedx}
\end{eqnarray}
where $k=25/448$ and $k^\prime=1/14$ are numerical constants, $G_F$ 
is the Fermi constant of weak decay, $E_\nu$ the neutrino energy.
From Eq.\ref{nudedx} the average fractional energy loss is
\begin{equation}
\frac{1}{E}\frac{dE}{dx}\frac{1}{\Gamma} = -\frac{k}{k^\prime} \approx 0.78
\end{equation}

\section{Energy cutoff}

Eq.\ref{nudedx} can be integrated, assuming $\delta_\nu$ independent from 
the energy, to obtain that a $\nu$ with initial energy $E_0$ after traveling a 
distance L will have energy E
\begin{equation}
E^{-5} - E_0^{-5} = 5k\delta_\nu^3 \frac{G_F^2}{192\pi^3}L \equiv E_T^{-5}
\label{et}
\end{equation}

$E_T$ acts as a spectral cutoff, therefore for any value of 
$\delta_\nu$ the high energy spectrum of a neutrino
beam would be depleted over a sufficiently long path.\\ 
The absence of such cutoff at ICARUS \cite{icarus} at LNGS ($L\approx 730\,\mathrm{km}$) up to
above $50\,\mathrm{GeV}$ and at NOMAD \cite{nomflu,nomlyu} ($L\approx 1\,\mathrm{km}$) 
above $200\,\mathrm{GeV}$ 
allows to set limits $\delta_\nu \leq 5\times 10^{-6}$.
Another effect that would deplete the $\nu$ distribution at high energy 
is the $\pi$ decay kinematics. The phase space for the decay $\pi\rightarrow \mu \nu_\mu$
for superluminal $\nu_\mu$ becomes smaller and smaller with increasing $\pi$ energy;
the effect is reducing the flux of high energy $\nu$ down to zero above some threshold
\cite{mestres,nussinov}.\\
As noted in \cite{glashow,bi}, that contradicts measurements of high energy neutrino
interactions in underground detectors.
The most stringent limits can be derived analyzing \cite{bi} high energy 
atmospheric $\nu_\nu$ ($\overline \nu_\mu$) detected in underground detectors \cite{icecube}.
\\
An alternative approach is to look for the production of isolated $e^+e^-$ pairs in
neutrino detectors traversed by a high energy, high intensity neutrino flux such as 
the NOMAD detector \cite{nomdet}.

\section{The NOMAD detector}

The NOMAD detector was designed for searching $\nu_\mu\rightarrow 
\nu_\tau$ oscillation at the CERN West Area Neutrino Facility (WANF) beam line.
The detector includes an active target of drift chambers (DC) with a 
mass of 2.7 tons and a volume $\approx 2.6 \times 2.6 \times \mathrm{4 m^3}$ 
complemented with electron 
identification provided by a Transition Radiation Detector 
(TRD) and an Electromagnetic Calorimeter (ECAL). 
This detector has proved effective in search of $\nu$ interaction with 
only one or few photons in the exclusive final state \cite{nomtauph,
nom339, nomscalar}.
The topology of these events is one or more e$^+$-e$^-$ pairs originating
in the target.

\section{The neutrino beam}

\begin{figure}
\onefigure[width=0.45\textwidth]{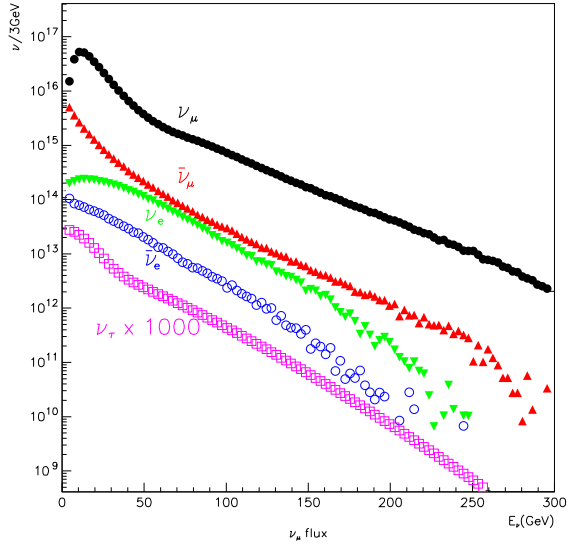}
\caption{Prediction of undistorted spectra of the different $\nu$ species crossing NOMAD 
($\nu_\tau$ is multiplied with 1000 for visibility) with the 1996-1998 data set
($4.1\,10^{19}$ p.o.t.)}
\label{neuspectra}
\end{figure}

The neutrino beam impinging on the NOMAD detector originates 
from the CERN West Area Neutrino Facility (WANF) \cite{wanf}; it is 
described in detail in \cite{nomflu}. The results are summarized in 
Fig.\ref{neuspectra} that shows the undistorted ($\delta_\nu=0$)
total neutrino fluxes, subdivided in the different components, traversing 
the active area of the NOMAD detector ($\approx 2.6 \times 2.6 \mathrm{m^2}$) 
during the period 1996-1998 for a total of $4.1\,10^{19}$ protons on target. 
The $\nu_\tau$ component, deduced from \cite{nutaunim}, is supposed to 
equal the $\overline \nu_\tau$ component.

\section{The $e^+e^-$ pair spectrum in NOMAD}
\begin{figure}
\onefigure[width=0.45\textwidth]{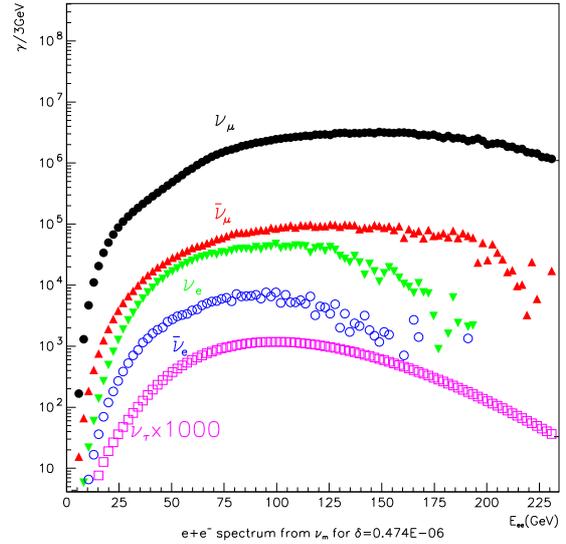}
\caption{Spectra of the $e^+e^-$ pairs produced by the different $\nu$ species 
in NOMAD ($\nu_\tau$
is multiplied with 1000 for visibility) with the 1996-1998 data set
($4.1\,10^{19}$ p.o.t.) accounting for spectral distortion
caused by $\nu$ energy loss and for $\delta_\nu = 0.5\times 10^{-6}$. 
An average detection efficiency $\epsilon=0.26$ is included.}
\label{gammaspectra}
\end{figure}

The spectral forms of the $e^+e^-$ pairs produced over a $4.0$ m length 
in the NOMAD detector are affected for $\delta_\nu \geq 10^{-6}$
by $\nu$ spectral distortion and for $\delta_\nu \leq 10^{-9}$ by threshold effects.\\
For intermediate values the spectra simply scale proportionally to $\delta_\nu^3$.
The different spectral components according to Eq.\ref{nudedx} are 
shown in Fig.\ref{gammaspectra} for $\delta_\nu=0.5\times 10^{-6}$, a
representative value of the scaling region. 
In order
to ease a comparison with existing experimental data, the fluxes have been multiplied
by an average detection efficiency $\epsilon=0.26$ deduced by \cite{nomtauph}.\\
The most appropriate comparison with existing NOMAD data is \cite{nomtauph}, where
$e^+e^+$ pairs from decay of a heavy neutrino mixing with $\nu_\tau$
were searched. This search established that the number of $e^+e^-$ pairs is compatible
with the expectations and additional sources can contribute no more than 
${\cal O}$(1) events.


\section{Limits on $\delta_\nu$}

\begin{figure}
\onefigure[width=0.45\textwidth]{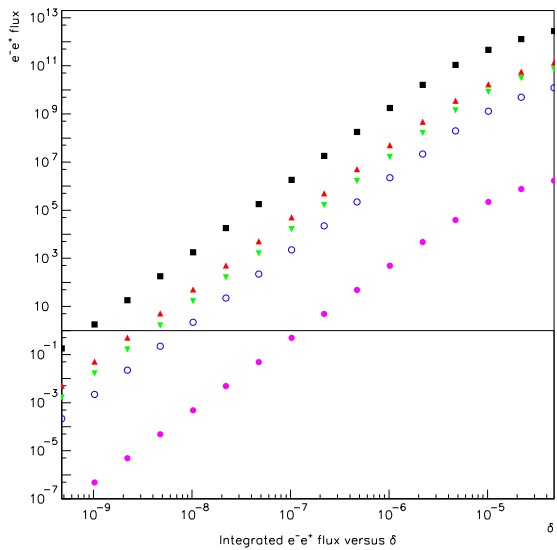}
\caption{$e^+e^-$ flux versus $\delta_\nu$ for the various $\nu$ species 
(colors and symbols are those used in Fig.\ref{gammaspectra}).
An average detection efficiency $\epsilon=0.26$ is included. The horizontal
line corresponding to 1 $e^+e^-$ pair defines the approximate upper limit.} 
\label{eevsdelta}
\end{figure}

The integrated $e^+e^-$ fluxes $F_{e^+e^-}$ ($\delta_\nu=0.5\times 10^{-6}$) 
for each species 
in the detector from Fig.\ref{gammaspectra} are in the second 
column of Tab.\ref{tabflux}, while in the last column there are the values 
of $\delta_\nu$ for which one $e^+e^-$ pair is predicted.
The last column is calculated accounting for distortion and 
threshold effects as shown in Fig.\ref{eevsdelta}, but it follows very 
closely the scaling law from
Eq.\ref{nudedx}, $\delta_\nu = 0.5\times 10^{-6}/F_{e^+e^-}^{1/3}$.\\
We emphasize that the limits derived with this analysis are approximate: 
the efficiency from \cite{nomtauph} is relative to a different (softer) $e^+e^-$ 
energy range, the spectrum of the process Eq.\ref{nuee}c is assumed monochromatic
and the statistical analysis is very crude. A dedicated analysis should be performed 
by the NOMAD collaboration to obtain more precise limits.\\
Existing limits are $\delta_{\nu_e}< 4.0\,10^{-9}$ from SN1987 \cite{longo1987} and 
$\delta_{\nu_{\mu}}< 1.4\,10^{-8}$ \cite{glashow}; in \cite{glashow} stronger limits 
from high energy events in IceCube are also presented for an unspecified $\nu$ species.
Recently, following an approach similar to that presented in this paper, the ICARUS 
collaboration set the limit $\delta< 2.5\,10^{-8}$ \cite{icarus}, 
presumely to be applied to $\nu_\mu$ .\\
The limit in Tab.\ref{tabflux} is the only one up to date on $\delta_{\nu_\tau}$.\\
Following \cite{amelino1} we remark that these limits are valid for the extensions of 
the special relativity theory with a so called 'broken' Lorentz invariance, for which the 
processes in Eq.\ref{nuee} take place; alternative extensions of the theory
with a so called 'deformed' Lorentz invariance, do not predict these processes and are 
not constrained by this analysis. 
Neither the possibility of a tachyonic superluminal neutrino is constrained 
by this analysis; nevertheless the discussions in \cite{amelino2,drago}
stress the difficulty of reconciling the data from accelerator experiments with this 
interpretation.

\begin{table}
\caption{Total $e^+e^-$ pairs expected for $\delta_\nu=0.5\times 10^{-6}$} for each species and 
$\delta_\nu$ giving one $e^+e^-$ pair.
\label{tabflux}
\begin{center}
\begin{tabular}{lll}
$\nu$ species  & $F_{e^+e^-}$ & $\delta_\nu$\\
$\nu_{\mu}$  & $1.8 \times 10^{8}$ & $8.4\, 10^{-10}$ \\
$\overline \nu_{\mu}$ & $5.1 \times 10^{6}$ & $2.7\, 10^{-9}$ \\
$\nu_e$  & $1.6 \times 10^{6}$ & $4.0\, 10^{-9}$ \\
$\overline \nu_e$ & $2.2 \times 10^{5}$ & $7.8\, 10^{-9}$ \\
$\nu_\tau (\overline \nu_\tau)$ & $4.8 \times 10^{1}$ & $1.3\, 10^{-7}$
\end{tabular}
\end{center}
\end{table}

\section{Conclusions}

We set strong bounds on special relativity violating processes involving neutrinos and 
anti-neutrinos of all species based on previous searches of isolated $e^+e^-$ pairs in 
the NOMAD detector. This translates in strong limits on possible superluminal behaviours
of neutrinos of all species for extensions of the special relativity theory with 'broken'
Lorentz invariance.\\
We strongly 
encourage the NOMAD collaboration to perform a dedicated analysis to optimize these limits.

\acknowledgments

Thanks to Prof. S.R. Mishra from University of South Carolina, Columbia, SC, USA and Prof.
L. Di Lella from CERN, Geneva, Switzerland for having supplied the neutrino spectra at NOMAD.

\bibliographystyle{eplbib}
\bibliography{NeutrinoNomad}

\end{document}